\documentclass{iau}

\usepackage{amsmath}
\usepackage{graphicx}
\usepackage{multirow}
\usepackage{jrnl_names}
\newcommand{\txw}{\textwidth}

\begin{document}

\lefttitle{Xie et al.}
\righttitle{Cycle variation of MFR axial orientations}

\jnlPage{1}{7}
\jnlDoiYr{2024}
\doival{10.1017/xxxxx}
\volno{388}
\pubYr{2024}
\journaltitle{Solar and Stellar Coronal Mass Ejections}

\aopheadtitle{Proceedings of the IAU Symposium}
\editors{N. Gopalswamy,  O. Malandraki, A. Vidotto \&  W. Manchester, eds.}

\title{Solar Cycle Variation of Axial Orientations and Favorable Locations of Eruptive MFRs}

\author{Hong Xie$^{1,2}$, Nat Gopalswamy$^1$, Sachiko Akiyama$^{1,2}$, Pertti Makela$^{1,2}$, and \\ Seiji Yashiro$^{1,2}$}
\affiliation{{}$^1$NASA Goddard Space Flight Center, 8800 Greenbelt Road Greenbelt, MD, 20771, USA \\email: \email{hong.xie@nasa.gov}}
\affiliation{{}$^2$The Catholic University of America, 620 Michigan Avenue, N.E. Washington, DC 20064, USA}

\begin{abstract}
Using multi-viewpoint observations from STEREO and SOHO during three solar cycles from 23 to 25, 
we study the magnetic flux rope (MFR) structures of coronal mass ejections (CMEs) near the Sun and magnetic clouds (MCs) at 1au. 
The study aims to investigate two phenomena: 1) the occurrence rate of CMEs near Hale sector boundaries (HBs) and 
2) solar-cycle variation of MFR axial orientations in CMEs and MCs. 
Our preliminary results include: 
1) the axes of MFRs in cycle 25 present a systematic northward orientation, which is 
the same as in cycle 23 but opposite to cycle 24;
2) the majority of the MFRs occurred near HBs (within 30 degrees) and some exceptional 
events occurred at non-HBs;
3) the axial fields in MCs present a similar north-south orientation, which changes from cycle to cycle. 
We discuss the implication of solar cycle variations of MFR axial orientations for space weather forecasts.
\end{abstract}


\maketitle

\section{Introduction}

Since the start of the previous century, solar observations have revealed that active regions (ARs) spot polarities present a systematic
east-west orientation that reverses in opposite hemispheres and across odd and  even cycles (Hale's law, \citealt{Hale25}). 
They also present a north-south orientation, i.e., the leading spots  appear closer to the equator than the following spots do and the tilt of
the spots with opposite polarities grows with the  latitude of AR to facilitate the polar field reversal in each cycle (Joy's law, \citealt{Hale19}).

Solar eruptions, such as flares and  coronal mass ejections (CMEs), are also influenced by large-scale neutral lines in
the solar corona, i.e., the boundaries between opposite magnetic poles \citep{Wilcox65}.  
\citet{Svalgaard75} showed that the sector boundary observed in magnetic fields at Earth can be mapped back to 
the large-scale neutral line in the photosphere and in the solar corona.  
\citet{Svalgaard76,Svalgaard11} found that magnetic field fluxes tend to emerge near the Hale sector boundary 
(HB hereafter), and suggested a link between the HB effect to the deep interior of the Sun. 
A number of other studies have also found that flares and most complex CME-productive ARs tend to occur near the HB 
\citep[e.g.][]{Getachew17, Gyenge17, Loumou18}.

The magnetic flux rope (MFR)  axial field in bipolar ARs and filament channels (FLs) can be approximated by the tilt of polarity
inversion lines (PILs) between opposite polarities. Recent studies have shown that most CMEs and magnetic clouds (MCs) contain the MFR structures
\citep[e.g.][]{Gopal13,  Xie13, Vourlidas14}; and the axial tilts of CMEs and MCs are well correlated with the PILs of their source regions 
\citep[e.g.][]{Marubashi15, Xie21}. Similar cycle variations have been found in MC axial tilts at 1au \citep[e.g.][]{Bothmer97,
Gopal2015, Lepping11}. In this paper, we investigate the solar-cycle variation of the MFR axial orientation and the relationship between HBs, 
PIL tilts and  CME eruptions.
The study aims to answer two questions: 1) how frequently or at what rate do CMEs erupt at HBs?  2) How does the 
solar-cycle variation of axial orientation of MFRs affect geomagnetic activity?

\section{Data and Event Selection} Seventy-seven MC-CMEs during the first four years of three cycles from 23 to 25 have been used in this
study. Events in cycles 23 and 24 are selected from \citet{Xie21}, where we have excluded complex events involving successive CMEs. 
The selected MC-CME pairs are divided into three types, i.e., stealth CMEs, CMEs
associated with flares (flare CMEs), and CMEs associated with filaments (filament CMEs).  
In Table~\ref{table_one}, we list the number of CMEs in each cycle and CME types.

\begin{table}[h!]
 \centering
 \caption{Numbers of MC-CMEs associated with three source types}\label{table_one}
 {\tablefont\begin{tabular}{@{\extracolsep{\fill}}lcrrr}
    \midrule
     & Total&Stealth CMEs& Flare CMEs& Filament CMEs\\
    \midrule
    SC 23& 26& 7&  11& 8\\
    SC 24& 27& 7&   9& 11\\
    SC 25& 24& 7&   8& 9\\
    \midrule
    \end{tabular}}
\end{table}

To identify the MFR sources and PIL tilts, we used extreme ultraviolet (EUV) images from 
the Atmospheric Imaging Assembly \citep[AIA;][]{Lemen12} 
and magnetograms from the Helioseismic and Magnetic Imager \citep[HMI;][]{Scherrer12} 
aboard the Solar Dynamics observatory \citep[SDO;][]{Pesnell12}.
For stealth CMEs, since the associated EUV and magnetogram signatures are too faint,
we apply a flux-rope fit to determine their source locations, we then use the coronal neutral 
line (CNL) tilt combined with the hemispheric helicity rule as a proxy for the PIL orientation \citep[see details in][]{Xie21}.  

\section{Results and Discussion}

\begin{figure}[ht]
  \centering 
  \includegraphics[scale=.9]{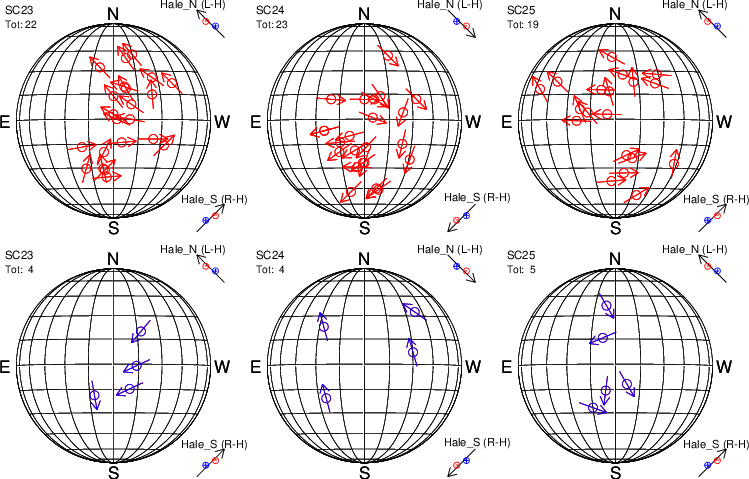}                                                                                                                                      
  \caption{Comparison of PIL tilts for Hale and non-Hale source regions during odd cycles (23 and 25) and even
cycle 24.}
  \label{Fig1:tilt}
 \end{figure}

Figure~\ref{Fig1:tilt} shows distributions of the MFR tilts for each cycle: (top) Hale CMEs and (bottom) non-Hale CMEs,
where  Hale (non-Hale) CMEs refer to cases when associated source polarities are following (against) the Hale's polarity
law.  We can see that, for all three cycles, the majority of the CMEs ($\sim$83\%) are Hale-CMEs.  
It is shown that during odd (even) cycles, the Hale MFR axial field has a systematic northward (southward) component. 
In cycles 23 and 25, the Hale
MFR axial directions in the northern hemisphere are northeast with a negative helicity (left-handed), and northwest in the
southern hemisphere with a positive helicity (right-handed).  In even cycle 24, the Hale MFR
axial tilts are opposite to the odd cycle, i.e., change to southwest and southeast directions.

\begin{figure}[ht]
  \centering
  \includegraphics[scale=.52]{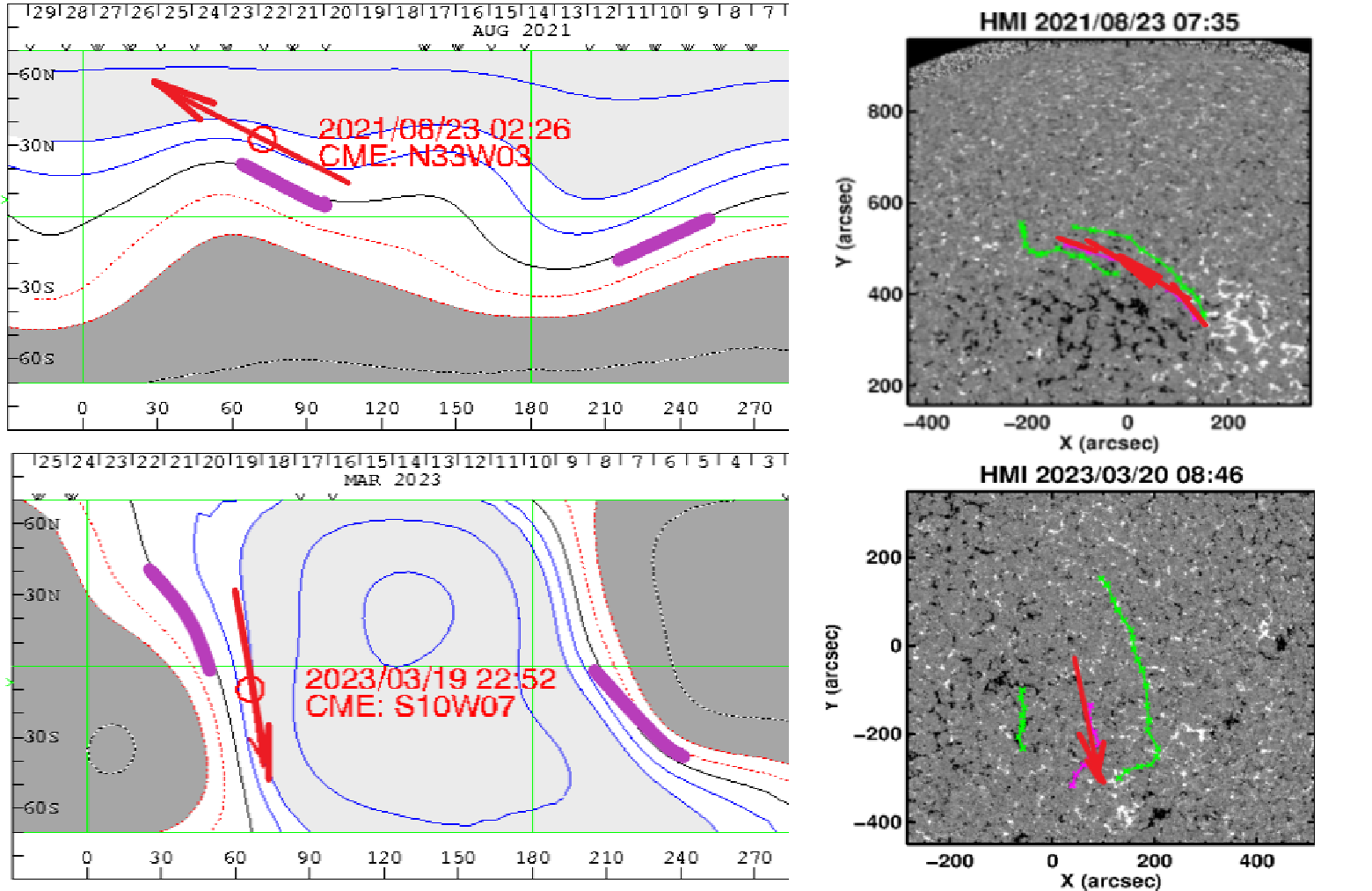}                                                                                                                                    
  \caption{(Left) WSO source surface synoptic maps calculated for a height of 2.5 Rs. Dark gray 
contours represent negative (-) fields, light gray contours, positive (+) fields. The black
solid line is the CNL. The CME source locations (red circles), PILs (red arrows) and HBs (thick purple lines) 
are overplotted on the maps. 
(Right) The CME PILs and post-eruption arcade footpoints (green lines) overplotted on the HMI magnetograms.}
  \label{Fig2:HBexmp}                                                                                                                                                      
\end{figure}

In Figure~\ref{Fig2:HBexmp}, we study the relative position of CME sources to the HB, where we overplot the CME source location
(red circle) on the Wilcox Solar Observatory (WSO) source surface map, and HBs are marked by thick purple lines.  
Here HB refers to the portion of a sector boundary where the polarity change across its neutral line matches that 
in bipolar source regions (ARs of FLs). 

Figure~\ref{Fig2:HBexmp} shows  (top) an example of a Hale MFR on 2021 Aug 23 and (bottom) a non-Hale MFR on 2023 Mar 19. 
In Figure~\ref{Fig2:HBexmp} top panels, the associated CME erupted near a northern HB, resulting in a northeast 
left-handed MC at 1 au, while in Figure~\ref{Fig2:HBexmp} bottom panels, a CME occurred near a southern non-HB, leading to a 
southward right-handed MC and well-known geomagnetic storm on 2023 Mar 24 during 
cycle 25 \citep[e.g.][]{Rajana24, Paul24}.  

By comparing the CME and HB locations, we found that $\sim$74\% of the MC-CMEs in our dataset
erupted near HBs (within $\pm$30$^\circ$); $\sim$26\% of CME-MCs occurred at the non-HBs,
suggesting that HB is one of the locations favorable to CME eruption.
The non-HB CMEs tend to occur at inter-Hale ARs (FLs) source regions 
and/or complex ARs accompanied with new emerging fluxes.

\begin{figure}[th]
    \centering
    \includegraphics[width=0.3\txw, height = 0.3\txw ]{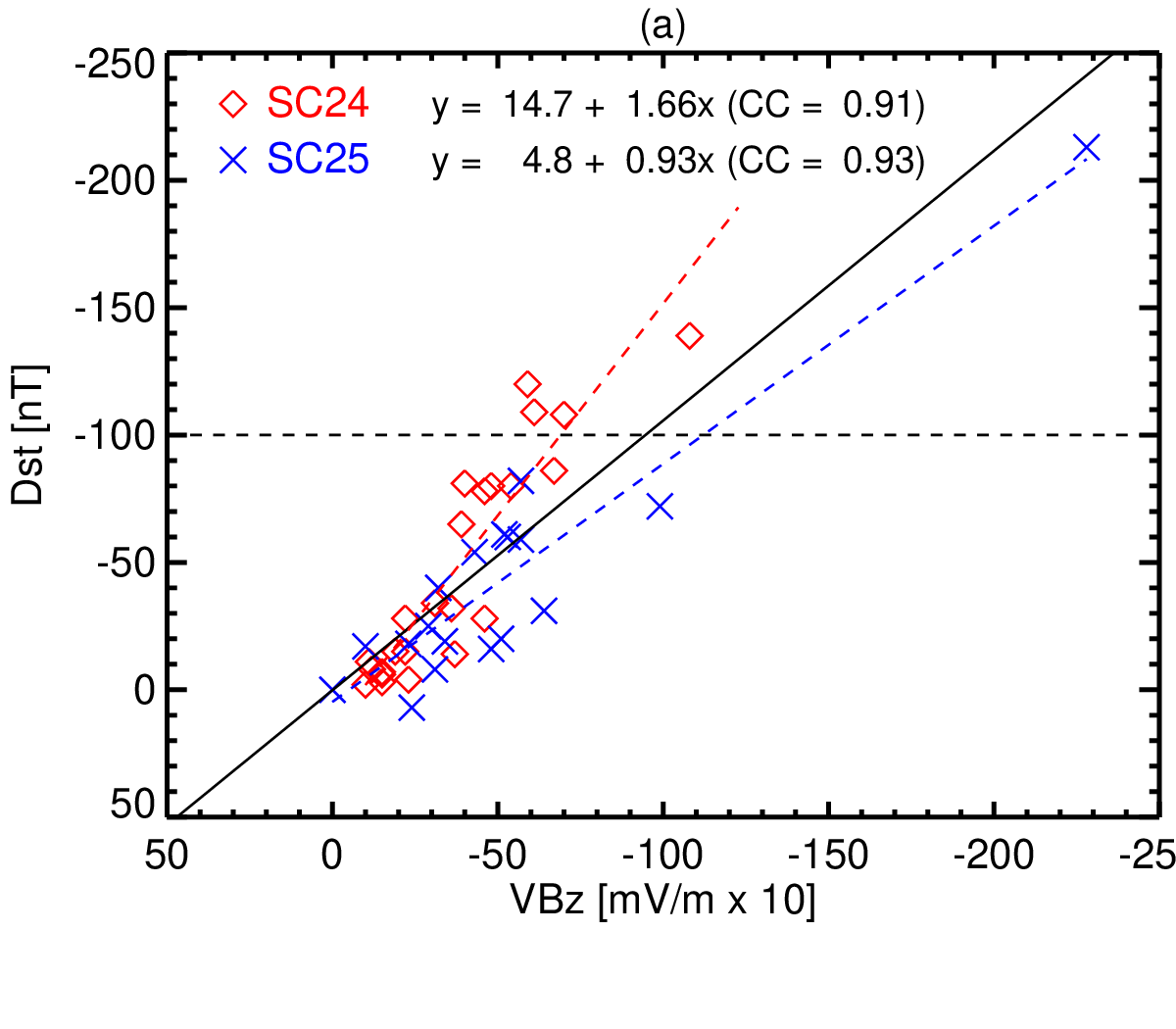}
    \includegraphics[width=0.3\txw, height = 0.3\txw ]{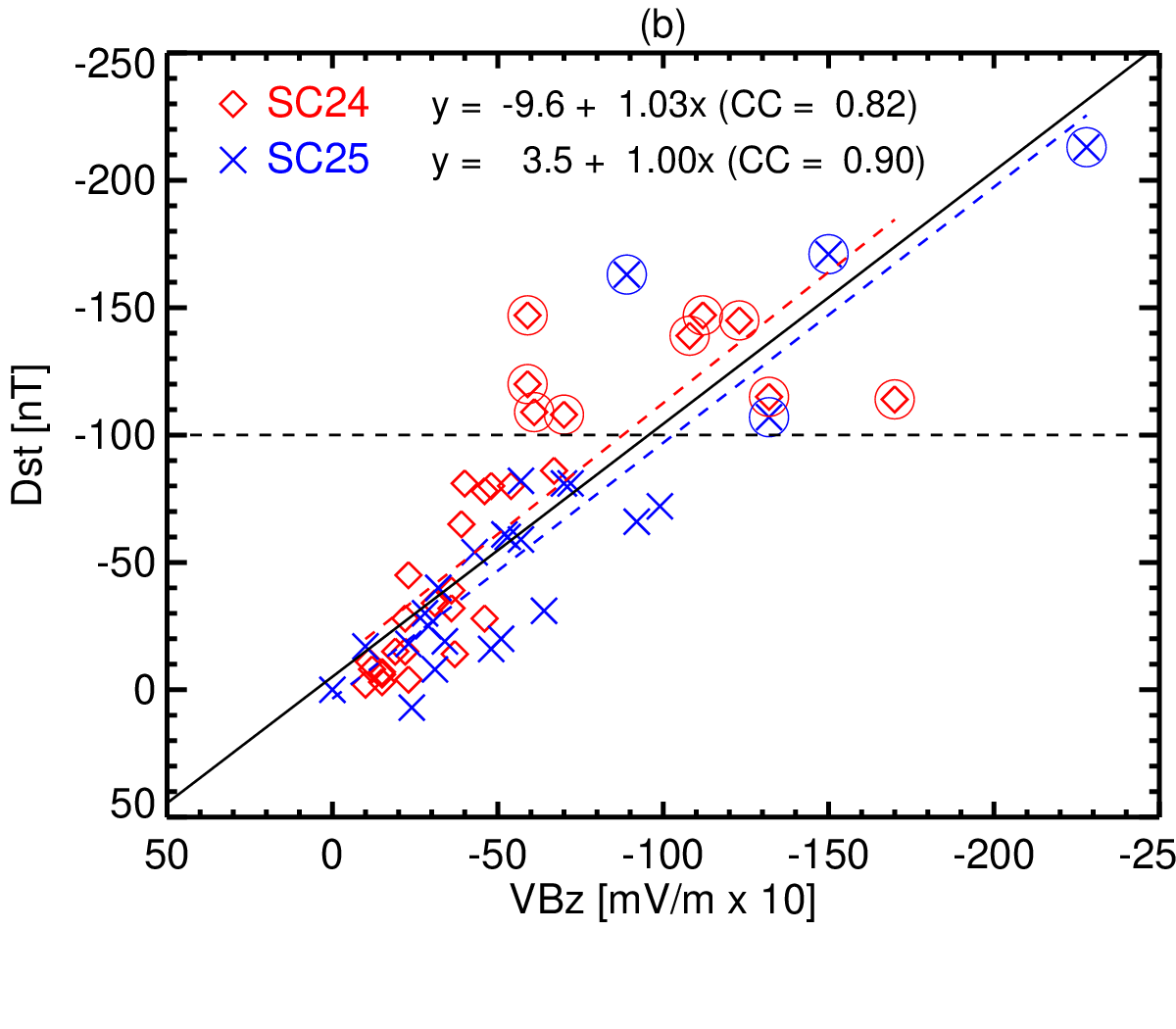}
    \includegraphics[width=0.3\txw, height = 0.3\txw ]{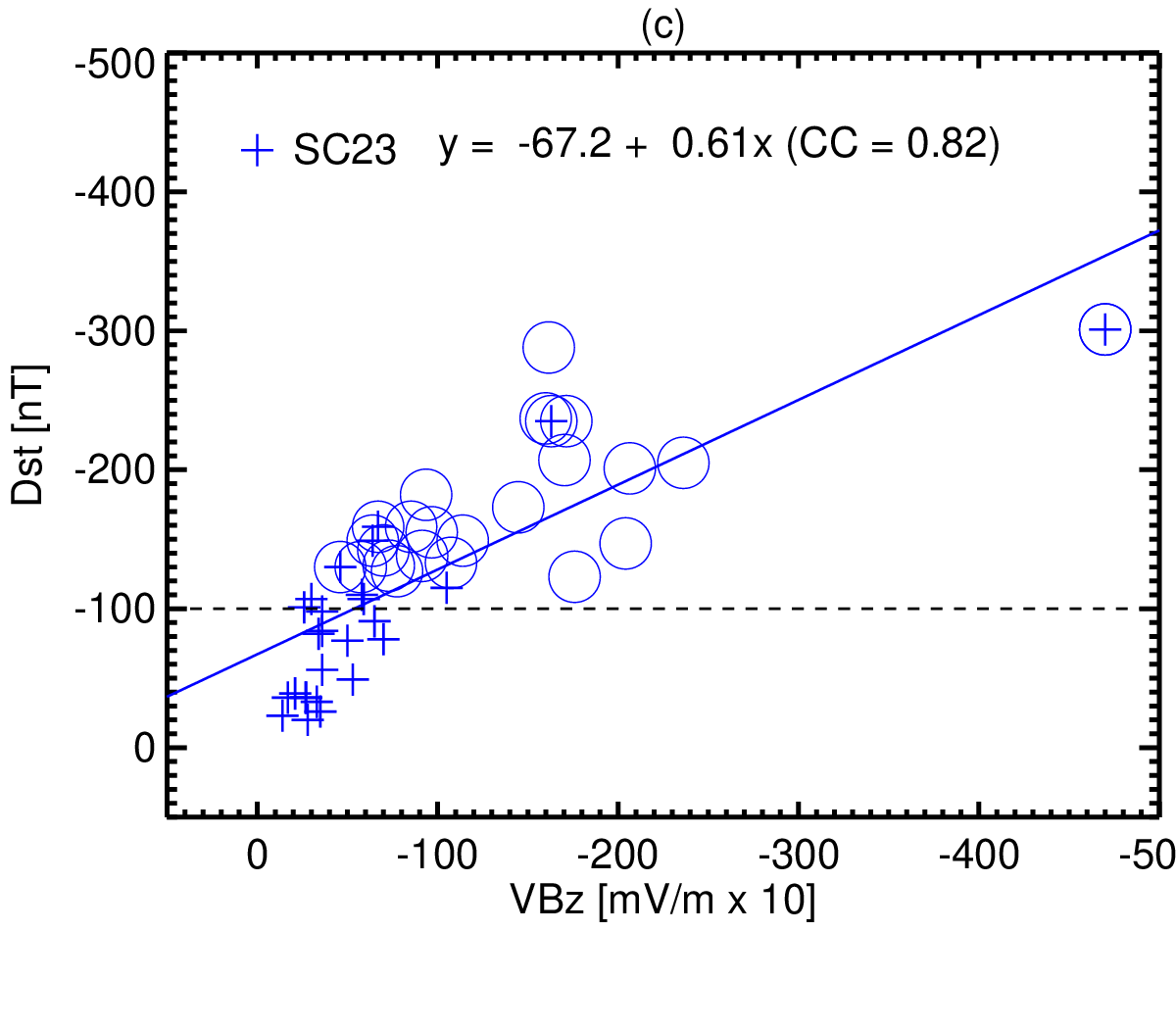}

   \caption{(a) Correlation between VB$_z$ and Dst during cycles 24 (red diamond) and 25 (blue 'x'), (b) same as (a), with major geomagnetic storms added,
marked by red and blue circles for the two cycles, and (c) correlation between VB$_z$ and Dst (blue cross) with major geomagnetic storms (blue circles) during cycle 23.}

  \label{Fig3:dst}                                                                                                                                                      
\end{figure}  

To discuss the effect of solar cycle variation of MFR axial orientations on geomagnetic activity,  in
Figure~\ref{Fig3:dst}a,  we compare the correlations between solar wind electric field (VB$_z$) and Dst index for Hale MC-CMEs during
cycle 24 (red diamond) and 25 (blue 'x').  
It shows that the fitted slope in cycle 24 is $\sim$1.8 times larger than in cycle 25,  i.e., 1.66 vs 0.93,   
suggesting that an even cycle with a prevailing southward axial field is more geoeffective than an odd cycle.
There are 9 major geomagnetic storms during the first four years of cycle
24 and only 4 major storms in cycle 25 from 2020 to 2023.
In Figure~\ref{Fig3:dst}b, we overplotted all the major geomagnetic storms with Dst $<$ -100 nT (marked by red and blue circles) for the
two cycles. 
Table 2 lists the number of major geomagnetic storms, where the storms are grouped as ejecta, sheath, Hale-MC and non-Hale MC storms, respectively. 
It is shown that the difference of fit-slopes for the two cycles has been smoothed out after adding ejecta, 
sheath and non-Hale storms.
During cycle 24, 4 out of 9 major storms are caused by the southward Hale MCs, 3 are associated with ejectas and 2 are sheath storms.
During cycle 25,  2 storms are caused by shock sheaths, 1 by a low-inclination MC (with a tilt$\sim$20$^\circ$), 1 by a southward non-Hale MC.
  
Note that, the geoeffective activity is not only affected by the MC axial orientation but also the solar cycle strength, i.e., 
the solar wind electric field.  
As shown in Figure~\ref{Fig3:dst}c, during an "active" cycle, the stronger cycle 23 has more number of major geomagnetic storms than both weaker cycles 24 and 25 due to 
the large VB$_z$. The geoeffectiveness during an acitve cycle may be enhanced due to interactions with other interplanetary magnetic
structures, primarily due to the compression of the existing B$_z$ magnetic field component  \citep[e.g.][]{Farrugia06, Lugaz17}.

\begin{table}[h!]
 \centering
 \caption{Numbers of different major geomagnetic storms during two cycles}\label{table_two}
 {\tablefont\begin{tabular}{@{\extracolsep{\fill}}lccrrr}
    \midrule
     & Total&\multicolumn{2}{c}{MC storms}& Ejecta storms& Sheath storms\\
     & &Hale & non-Hale& & \\

    \midrule
    SC 24& 9& 4 &0&   3& 2\\
    SC 25& 4& 2&1&0& 2\\
    \midrule
    \end{tabular}}
\end{table}

\section{Conclusions} 

1) The MFR PILs of cycle 25 present a systematic northward orientation as in cycle 23 but opposite to cycle 24
(southward orientation).  This cycle variation may explain that, although cycle 25 is slightly stronger, it contains a
relatively weaker geomagnetic activity than cycle 24 during its first four years. 
2) $\sim$74\% CME-MCs erupted near HBs (within 30$^\circ$), suggesting HB is one of the locations favorable to CME eruption.

We thank SOHO, STEREO, and SDO teams for making their data available online. 
H.X., S.A., and S.Y. are partially supported by NSF grant AGS-2228967. P.M. is
partially supported by NSF grant AGS-2043131.

\bibliography{iau}
\bibliographystyle{aasjournal}
\end{document}